\documentclass[aps,prb,twocolumn,showpacs,showkeys,superscriptaddress,10pt]{revtex4}

\usepackage{amsmath}
\usepackage{amssymb}
\usepackage{latexsym}
\usepackage{indentfirst}
\usepackage{graphicx}
\usepackage{color}
\usepackage{bm}
\usepackage{multirow}

\begin{document}

\title{Temperature dependence of the switching field distributions in all-perpendicular spin-valve nanopillars}

\author{D.~B. Gopman}
\email{daniel.gopman@physics.nyu.edu}
\affiliation{Department of Physics, New York University, New
             York, NY 10003, USA}
\author{D. Bedau}
\affiliation{Department of Physics, New York University, New
             York, NY 10003, USA}
\affiliation{HGST San Jose Research Center,
             San Jose, CA 95135 USA}
\author{G. Wolf}
\affiliation{Department of Physics, New York University, New
             York, NY 10003, USA}
\author{S. Mangin}
\affiliation{Institute Jean Lamour, UMR CNRS 7198, Nancy Universit\'{e},
             Vandoeuvre, France}
\author{E.~E. Fullerton}
\affiliation{CMRR, University of California at San Diego,
             La Jolla, CA 92093, USA}
\author{J.~A. Katine}
\affiliation{HGST San Jose Research Center,
             San Jose, CA 95135 USA}
\author{A.~D. Kent}
\affiliation{Department of Physics, New York University, New
             York, NY 10003, USA}
\begin{abstract}
We present temperature dependent switching measurements of the Co/Ni multilayered free element of 75~nm diameter spin-valve nanopillars. Angular dependent hysteresis measurements as well as switching field measurements taken at low temperature are in agreement with a model of thermal activation over a perpendicular anisotropy barrier. However, the statistics of switching (mean switching field and switching variance) from 20~K up to 400~K are in disagreement with a N\'{e}el-Brown model that assumes a temperature independent barrier height and anisotropy field. We introduce a modified N\'{e}el-Brown model thats fit the experimental data in which we take a $T^{3/2}$ dependence to the barrier height and the anisotropy field due to the temperature dependent magnetization and anisotropy energy. 
\end{abstract}


\maketitle

In nanometer scale magnetic elements, the N\'{e}el-Brown model is widely invoked due to its reducing the dynamics of a large assembly of spins to the behavior of a single, collective spin orientation in a uniaxial potential energy landscape \cite{wB63,lN49,dA95}. From this model there have emerged important predictions for thermally assisted reversal under applied magnetic fields and spin-torques \cite{eS48,zL04}. 

Recent spin-torque and thermally assisted switching studies in perpendicularly magnetized nanopillar spin-valves have tested this macrospin model at room temperature. Experimentally obtained energy barrier heights were shown to be much lower than the uniaxial barrier height determined by the entire macrospin volume \cite{jS02,dG12,dB10}. Nevertheless, the switching distributions appear well described by overcoming a single energy barrier, whose height is related to an excited magnetic subvolume in the free layer element \cite{jS11}. This description is in good agreement with spin-torque switching results on 100~nm lateral size nanopillar devices and the underlying behavior has been observed in micromagnetic simulations as well as dynamic imaging measurements \cite{dB11}.

In order to test the validity of this uniaxial model, temperature dependent measurements of the switching statistics can be used to probe the barrier height. We present dynamical measurements of the switching field in Co/Ni nanopillars as a function of temperature. While our switching field measurements taken at constant temperature are consistent with a single energy barrier model, the temperature dependence of the mean switching field and switching distribution width suggest that the barrier is not independent of temperature. This can be reconciled with experimental data by taking account of the temperature dependence of the magnetization and the perpendicular anisotropy previously noted in Co/Ni multilayered films, which modify the barrier height and coercivity. By introducing a $T^{3/2}$ dependence to the barrier height and coercivity to the N\'{e}el-Brown model, we demonstrate the validity of this model for thermally activated reversal over a wide temperature range (50~K - 400~K). 

The Co/Ni nanopillars studied here are part of an all perpendicular spin-valve device. Details on materials and sample preparation have been reported previously \cite{sM06}. The magnetic multilayered structure consists of a Pt(3 nm)/[Co(0.25   nm)/Pt(0.52   nm)]x5/Co(0.2   nm)/[Ni(0.6   nm)/ Co(0.1 nm)]x2/Co(0.1 nm) hard reference layer and a Co(0.1 nm)/[Co(0.1 nm)/Ni(0.6 nm)]x4/Pt(3 nm) free layer separated by  a  4  nm  Cu  spacer  layer patterned into 75~nm diameter nanopillars. Measurements were taken in a closed-cycle cryostat between the poles of an electromagnet oriented perpendicular to the device plane and at temperatures ranging from 20~K - 400~K. The reference layer magnetization switches for an applied field close to 1 T. Since no fields greater than 0.5 T are applied during the measurements, the reference layer is expected to remain stable. For all the experiments shown here the reference layer magnetization points along the direction of positive magnetic fields. 

\begin{figure}[htb]
  \begin{center}
   \includegraphics[width=3.0in,
    keepaspectratio=True]
   {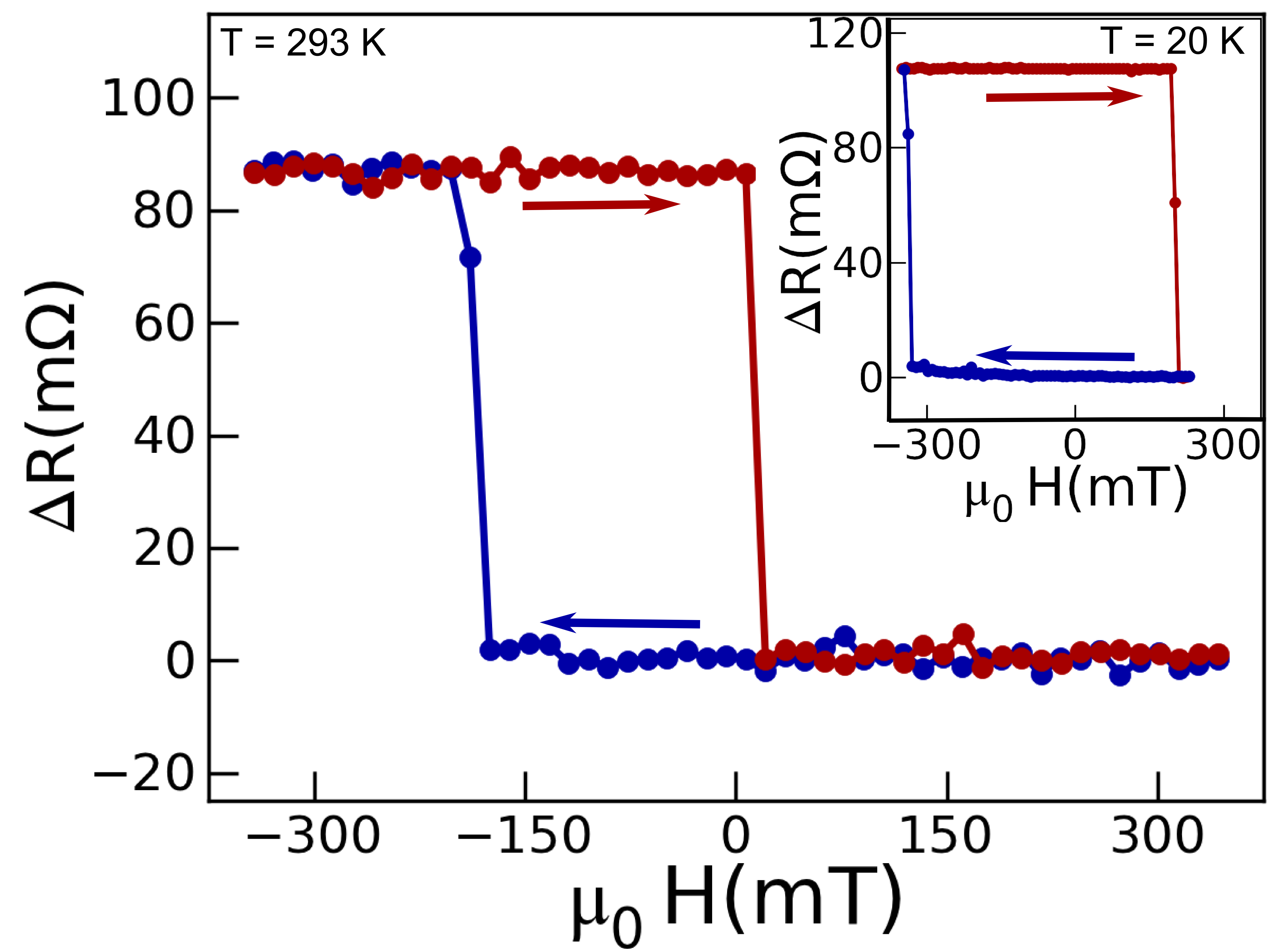}
  \end{center}
  \caption{\label{fig:Hysteresis} Quasistatic resistance versus applied perpendicular field hysteresis loop of a 75 nm diameter Co/Ni device at room temperature. $\mathrm{R = 7.5 \, \Omega}$. Inset: GMR Hysteresis Loop at T=20~K. $\mathrm{R = 5.9  \, \Omega}$.}
\end{figure}

The magnetization of the free layer is probed indirectly with four-probe measurements of the differential resistance of the spin-valve device under an $50 \, \mu A$ excitation current using standard lock-in techniques.  Figure~\ref{fig:Hysteresis} shows typical resistance versus applied perpendicular field hysteresis loops at room temperature and at 20~K. The sharp changes in resistance $\Delta R$ referenced to the low resistance parallel alignment indicates switching of the free layer into a parallel or antiparallel configuration with the reference layer. The approximately 80~mT shift of the center of the hysteresis loops reflects the inhomogeneous dipolar field from the reference layer averaged over the free layer. This spatially inhomogeneous field can lead to asymmetric reversal behavior for $\mathrm{AP \rightarrow P}$ and $\mathrm{P \rightarrow AP}$ transitions \cite{dG12}. In the following results, we will simply present the total effective field $H = H_{app} - H_D$, as the applied field minus the dipolar loop shift.

\begin{figure}[!b]
  \begin{center}
   \includegraphics[width=3.0in,
    keepaspectratio=True]
   {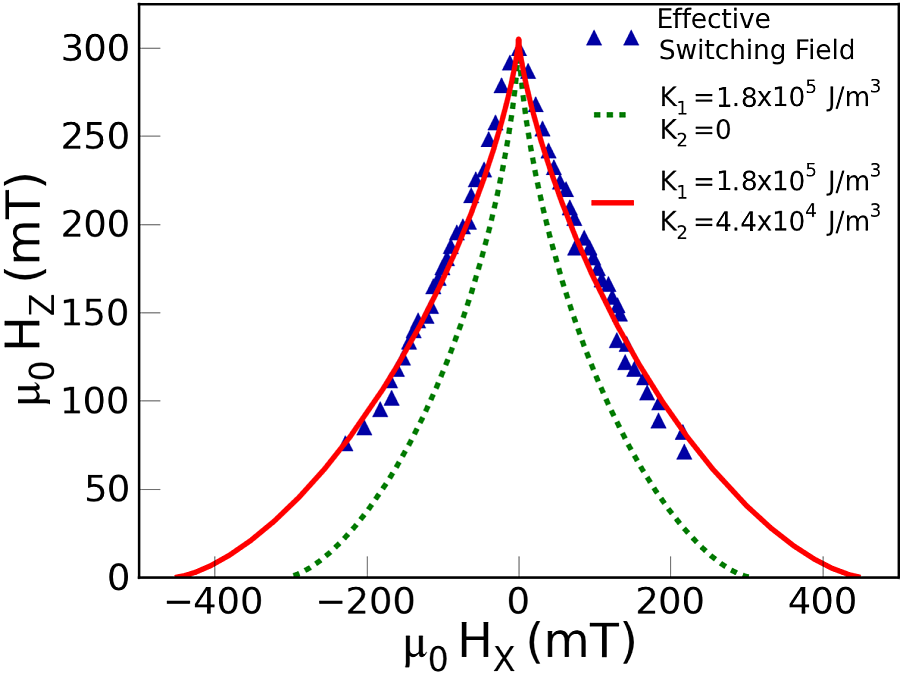}
  \end{center}
  \caption{\label{fig:SW Plot} Angular dependence of the switching field of a 75 nm diameter Co/Ni device at T=20~K. The blue triangles are measured switching field data. The hashed green line is the prediction of the Stoner-Wohlfarth model with only a first order uniaxial contribution to the anisotropy energy and the solid red line includes a second-order uniaxial anisotropy energy contribution.}
\end{figure}

In order to probe the switching behavior, we examine the angular dependence of reversal. Figure~\ref{fig:SW Plot} illustrates the switching fields measured at 20~K. These were determined from differential resistance versus field minor loops taken under different rotation angles of the spin-valve pillar with respect to the fixed poles of our electromagnet. The switching fields form an astroid that is clearly symmetric, but exhibits much larger hard-axis switching fields than predicted by the Stoner-Wohlfarth model. However, with the addition of a second-order uniaxial anisotropy contribution to the energy density we demonstrate a qualitatively better fit to our angular dependent data. The anisotropy energy density for our free layer element is therefore: \begin{eqnarray}
&& E(\theta) = K_1 \sin ^2 \theta + K_2 \sin ^4 \theta.
\label{Generalized_SW}
\end{eqnarray}
From the best-fit curve to our data, we obtain $K_1 \, = \, 9 \times 10^4 \, \mathrm{J/m^3}$ and $K_2 \, = \, 2 \times 10^4 \, \mathrm{J/m^3}$, assuming that $M_S \, = \, 600 \, \mathrm{kA/m}$. The presence of second-order anisotropies in Co/Ni multilayered films has been previously investigated using ferromagnetic resonance methods and their presence in spin-valve devices has been predicted as the source of symmetry breaking in the current-field state diagram \cite{jB07,yH09,sL12}. This result is a strong indication of the significance of the high uniaxial barrier to reversal and the extracted anisotropy field $\mu_0 H_K = 305 \, \mathrm{mT}$ will be used subsequently in expressions for the energy barrier. 

Thermally-assisted escape of a nanomagnet's magnetization from a metastable state can be described by an Arrhenius law in the following formalism: $ \Gamma = \Gamma _0 \exp \left( - \frac{ E ( H ) } { k_B T} \right)$, where $k_B$ is Boltzmann's constant, $T$ is the temperature, $\Gamma _0$ is in the range of 1-10~GHz and $E(H)$ is a field-dependent energy barrier. From this escape rate, we have the survival probability of a metastable state after time $t$: $P(t) = \exp \left ( - \Gamma t \right)$. The above expression assumes a model system described by a single field-dependent energy barrier, for which we use the following expression:\begin{eqnarray}
&& E(H) = E_0 \left(1 - H/H_{c0} \right)^\eta = E_0 \varepsilon ^ \eta
\label{Energy}
\end{eqnarray}
The expression describes the energy barrier in terms of a barrier height, $E_0$, the zero-temperature anisotropy field, $H_{c0}$, and $\eta$, an exponent that rapidly decreases from 2 to 1.5 under a small misalignment between the external magnetic field and the easy axis of the ferromagnet; we will take 1.5 as the value for $\eta$ \cite{rV89,wC95}. 

To test the single barrier model, we have conducted switching field measurements as a function of temperature. Switching field measurements constitute applying a linear magnetic field sweep and recording the field at which the magnet reverses, which for our spin-valve pillars is defined by a step change in device resistance as seen in Fig.~\ref{fig:Hysteresis} \cite{wW97,jS02}. Thermal activation predicts a characteristic distribution of switching fields sensitive to the sweeping rate and the temperature. For modeling our switching field measurements, we follow the change of variable introduced by Kurkij\"{a}rvi to transform the survival time expression into a survival probability versus field under a linearly ramped magnetic field ($v = dH/dt = const.$). Taking the derivative of the probability with respect to time, we have: $dP/dt = - \Gamma \exp \left ( - \Gamma t \right) = - \Gamma P$ \cite{jK72}. Rearranging terms, we have $\frac{dP}{P} = - \Gamma dt$, from which we apply the change of variable, $dt = dH / v$ and obtain $\frac{dP}{P} = - \frac{1}{v} \Gamma dH$. Finally, we integrate this expression to get this final expression of the survival probability as a function of field: \begin{eqnarray}
&& P_{NS}(H) = \exp \left[ -\frac{\Gamma_0}{v} \int^H_0 \exp \left[ \frac{-E(H')}{k_BT} \right ] dH' \right ]
\label{SFD}
\end{eqnarray}
Correspondingly, the switching probability as a function of field is given in terms of the derivative of the survival probability $p(H) = - \frac{dP}{dH}$.

\begin{figure}[!b]
  \begin{center}
   \includegraphics[width=3.0in,
    keepaspectratio=True]
   {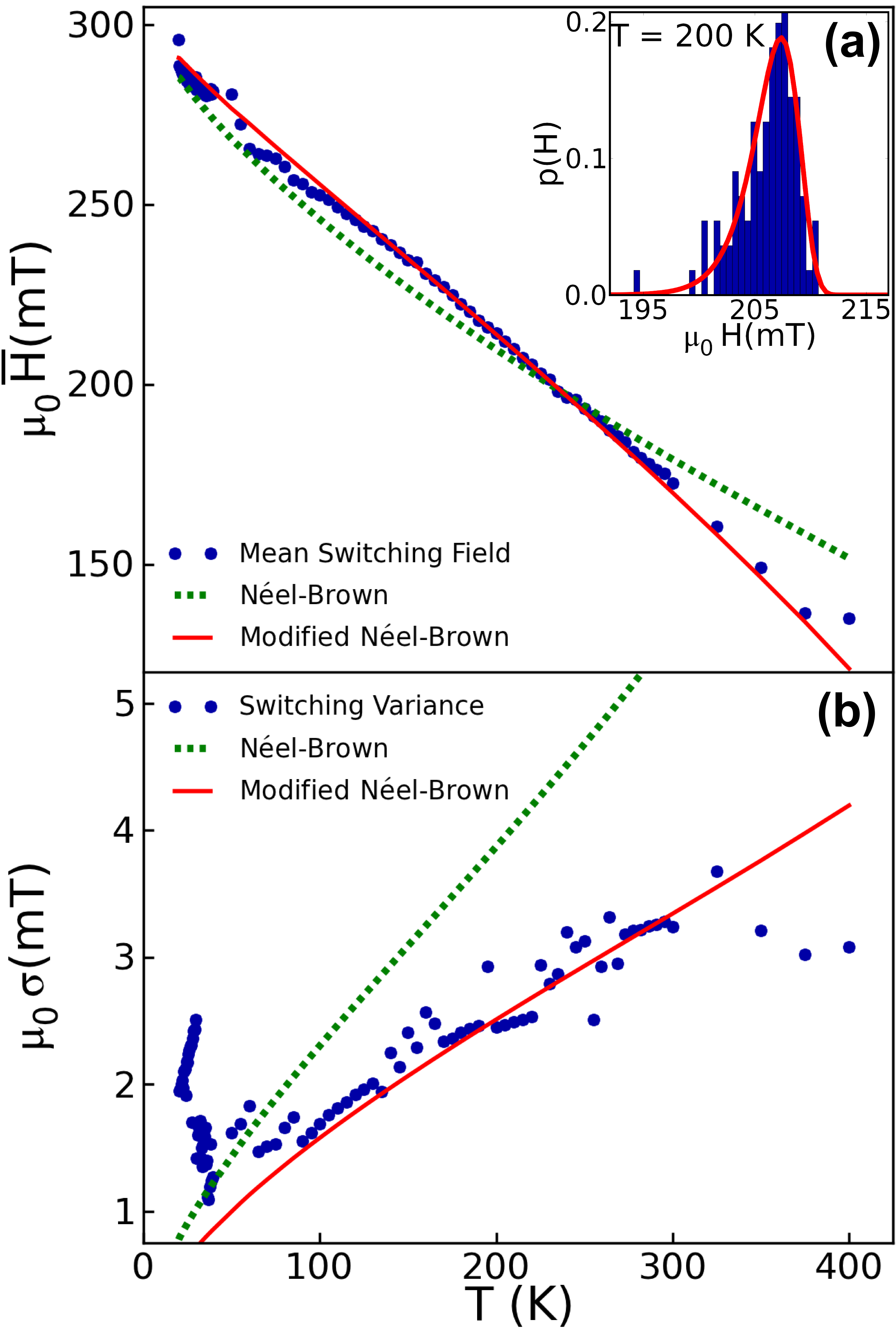}
  \end{center}
  \caption{\label{fig:Means_vs_Temps} : (a) Mean switching field versus temperature for a 75 nm diameter Co/Ni device. The dashed green line represents the best-fit assuming a temperature-independent barrier height, while the solid red line considers a temperature dependent energy barrier. Inset: Switching field histogram at 200 K. Histograms as a function of temperature are used to determine the mean and variance of the switching field. (b) Variance versus temperature [the dashed green line and solid red lines reflect the fitting parameters used in (a)]. }
\end{figure}

For each temperature we conducted 100 switching field measurements under a field sweeping rate of $50 \, \mathrm{mT/s}$, from which we have determined $\mu_0 \overline{H}$, the mean switching field, and $\sigma$, the variance of the switching field distribution. Figure~\ref{fig:Means_vs_Temps} displays the temperature dependence of the mean switching field and switching variance from 20~K to 400~K. According to our thermal activation model, the mean switching field and switching variance should follow the following expressions valid in the high barrier limit ($E_0 / k_B T \gg 1$) \cite{lG94,aG95}: \begin{eqnarray}
&& \overline{H} \cong H_{c0} \left( 1 - \left[ \xi \ln \left( \frac{ \Gamma _0 H_{c0} }{ \eta v \xi \varepsilon ^{ \eta - 1} } \right)  \right]^{1 / \eta} \right), \\
&& \sigma \cong H_{c0} \frac{1}{\eta} \left(\frac{1}{\xi} \right)^{1 / \eta} \left[ \ln \left( \frac{ \Gamma _0 H_{c0} }{ \eta v \xi \varepsilon ^{ \eta - 1} } \right)  \right]^{(1 - \eta) / \eta}, \\
&& \xi = E_0 / k_B T
\label{sfdd}
\end{eqnarray}
Assuming an attempt frequency of $\mathrm{\Gamma _ 0 = 1 \, GHz}$ and an exponent $\eta = 1.5$, for a sweeping rate of $v = 50 \, \mathrm{mT/s}$ and an experimentally determined anisotropy field ($\mu_0 H_{c0} = \, 305 \, \mathrm{mT}$) from Fig.~\ref{fig:SW Plot}, we obtain a best-fit to the mean switching field for a barrier height of $E_0/k_B = 20700 \, \mathrm{K}$. It is clear qualitatively from Fig.~\ref{fig:Means_vs_Temps}(a) that the mean switching field data is poorly fit by the green hashed line, and that Fig.~\ref{fig:Means_vs_Temps}(b) definitely does not agree with this best-fit parameter. 

Recent results on the saturation magnetization and anisotropy energy of similar Co-Ni multilayered films demonstrated a strong temperature dependence and low Curie temperature (435~K) \cite{oP09}. It has been demonstrated that the Bloch law temperature dependence for saturation magnetization provides a good description for magnetic thin films \cite{mP91,gL88}. Furthermore, \textit{ab initio} calculations suggest that the temperature dependence of uniaxial perpendicular anisotropy energy in single crystals and in sputtered films scales with the magnetization squared ($M^2$) \cite{jS06,aB07}. Equation~\ref{Energy} for the energy barrier therefore may have implicit dependence on the magnetization and anisotropy:
\begin{eqnarray}
&& E_0 = K(T)V - \frac{1}{2} \mu_0 M^2_S(T) V, \\
&& H_{c0} = \frac{2 K(T)}{M_S(T)} - \mu_0 M_S(T), \\
&& K(T) = K(0) \left(\frac{M_S(T)}{M_S(0)}\right)^2
\label{TempDepend}
\end{eqnarray}
where $V$ is the activation volume, $M_S$ is the saturation magnetization, and $K$ is the uniaxial anisotropy energy. We attribute a $T^{3/2}$ dependence of the magnetization due to the Bloch law temperature dependence of the magnetization and anisotropy field and neglect higher order terms in $T$, which is sufficient to fit the entire dataset: \begin{eqnarray}
&& E_B(T) \sim E_{0} \left(1 - 2 B_0 T^{3/2} \right), \\
&& H_{c}(T) = H_{c0} \left(1 - B_0 T^{3/2} \right),
\label{Bloch}
\end{eqnarray}
where $B_0 = 2.5 \times 10^{-5} K^{-3/2}$ was determined as a best-fit parameter for both the mean switching field and switching variance in Fig.~\ref{fig:Means_vs_Temps} and is comparable to the prior result on the temperature dependence of the saturation magnetization in a Co-Ni film \cite{oP09}. The solid red lines in Figs.~\ref{fig:Means_vs_Temps}(a)\&(b) demonstrate the best-fit mean switching field and switching variance trendlines as a function of temperature. We note however, that the increased variance at low temperatures (e.g. below 50~K) shown in Fig.~\ref{fig:Means_vs_Temps}(a) cannot be explained within the N\'{e}el-Brown model and is the subject of ongoing investigation. In the inset of Fig.~\ref{fig:Means_vs_Temps}(a), we also demonstrate the fit of our model switching probability density curve to our switching field distribution at 200~K. Given our anisotropy field value of 305~mT, we extrapolate a zero-temperature barrier height of $E_0/k_B = 35000 \, \mathrm{K}$. From these values, we can determine a thermally activated subvolume of $d \cong 48 \, \mathrm{nm}$, or approximately $41 \% $ of the estimated free layer volume, assuming a zero-temperature saturation magnetization of $M_S \cong \, \mathrm{600 \, kA / m}$.

\begin{figure}[!t]
  \begin{center}
   \includegraphics[width=3.0in,
    keepaspectratio=True]
   {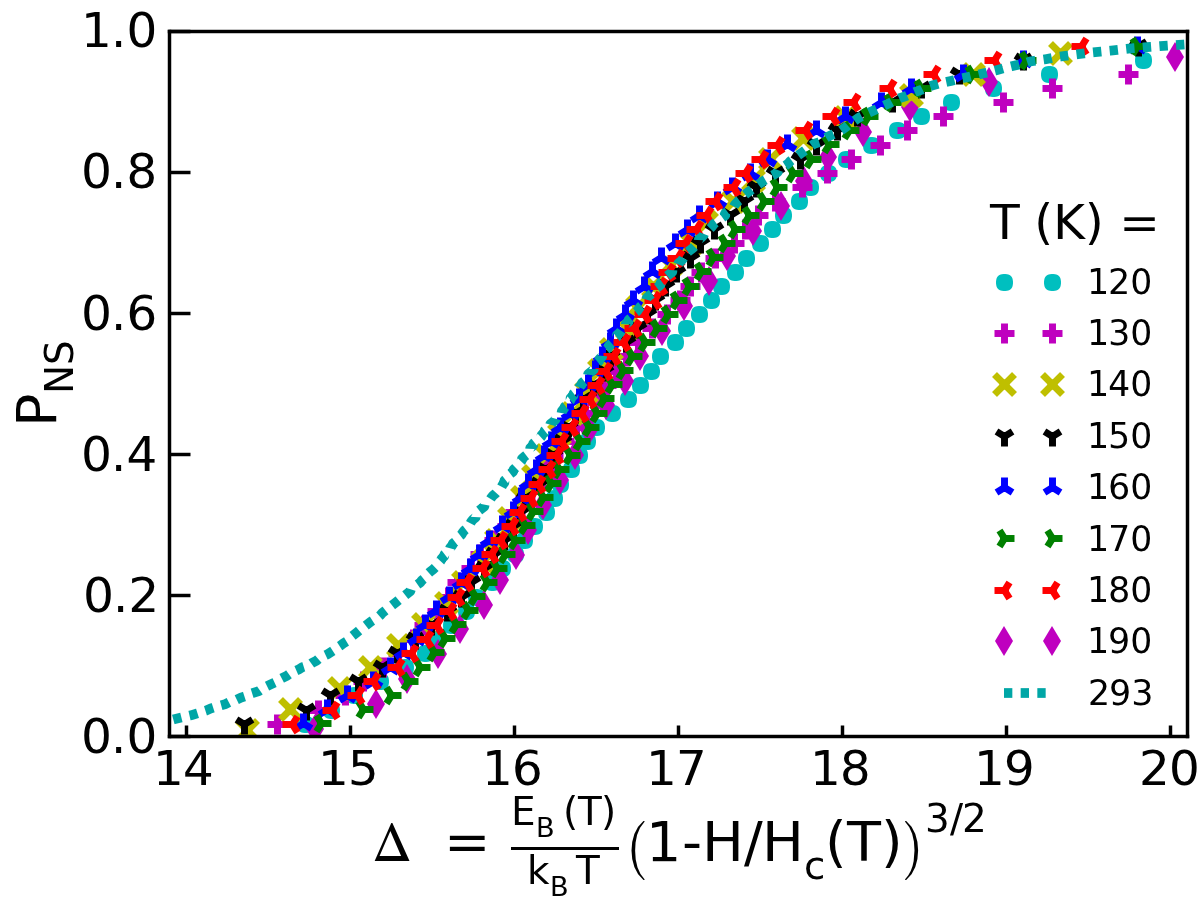}
  \end{center}
  \caption{\label{fig:UniversalCurve} Scaling plot of the survival function $\mathrm{P_{NS}}$ for temperatures between 120-300~K.}
\end{figure}

We present additional confirmation of this model on a second nanopillar device in which we acquired statistics from a greater number of switching events, large enough to fit the entire switching field distribution, in addition to the first moments. Figure~\ref{fig:UniversalCurve} displays a scaling plot of the survival function obtained from 1000 switching field measurements at temperatures ranging from 120~K to room temperature. In accordance with Equation~\ref{SFD}, the distributions collapse onto a master curve defined by the survival function $P_{NS}$, where we have the following expression for the energy barrier normalized to $k_B T$: \begin{eqnarray}
&& \Delta = \frac{E_B(T)}{k_B T} \left(1 - H / H_{c}(T) \right)^{3/2}.
\label{Delta}
\end{eqnarray}
Assuming the same temperature dependences for $E_B$, $H_c$ as in Eq.~\ref{Bloch}, we obtain $E_0/k_B = 41000 \, \mathrm{K}$ and 306~mT for the zero-temperature barrier height and anisotropy field, respectively. From these values, we determine a thermally activated subvolume of $d \cong 52 \, \mathrm{nm}$, or approximately $48 \% $ of the estimated free layer volume, again assuming a zero-temperature saturation magnetization of $M_S \cong \, \mathrm{600 \, kA / m}$.

In conclusion, we have demonstrated the sensitivity of the temperature dependent switching field measurements to the temperature-dependent material properties in Co/Ni multilayers. The data is consistent with a single energy barrier process described within the N\'{e}el-Brown model of magnetization reversal. Upon introducing a temperature dependence to the energy barrier, we demonstrate that the temperature evolution (50-400~K) of the mean switching field, switching variance and the survival function of Co/Ni multilayered nanopillars can be described by thermal activation over a single perpendicular anisotropy barrier. The agreement of our experimental data with this simple extension of the N\'{e}el-Brown model of magnetization reversal is also evidence of the temperature dependence of perpendicular anisotropy and magnetization in Co/Ni multilayered films. 



\section*{Acknowledgments}
This research was supported at NYU by NSF Grant Nos. DMR-1006575 and NSF-DMR-1309202, as well as the Partner University Fund (PUF) of the Embassy of France. Work at UCSD supported by NSF Grant No. DMR-1008654. 

\bibliographystyle{apsrev}

\end{document}